\documentclass[AMA,STIX1COL]{WileyNJD-v2}

\usepackage{graphicx}
\usepackage{tabularx}
\usepackage{setspace}
\usepackage{subcaption}
\usepackage{graphicx}
\usepackage{multirow} 
\usepackage{array}
\newcolumntype{C}[1]{>{\centering\arraybackslash}p{#1}}
\newcolumntype{L}[1]{>{\raggedright\let\newline\\\arraybackslash}p{#1}}
\newcolumntype{R}[1]{>{\raggedleft\let\newline\\\arraybackslash}p{#1}}

\articletype{Article Type}%

\received{\today}
\revised{----}
\accepted{----}

\usepackage{tikz}
\usepackage[printwatermark]{xwatermark}
\newcommand{\placetextbox}[3]{
  \setbox0=\hbox{#3}
  \AddToShipoutPictureFG*{
    \put(\LenToUnit{#1\paperwidth},\LenToUnit{#2\paperheight}){\vtop{{\null}\makebox[0pt][c]{#3}}}%
  }%
}%

\newsavebox\mybox
\savebox\mybox{\tikz[color=red,opacity=0.25]\node{PRE-PRINT};}
\newwatermark*[
  allpages,
  angle=45,
  scale=10,
  xpos=-30,
  ypos=20
]{\usebox\mybox}

\begin{document}

\title{Evaluation of the performance challenges in automatic traffic report generation with huge data volumes}

\placetextbox{0.58}{0.94}{\parbox{0.85\textwidth}{ \footnotesize{
    This is the pre-peer reviewed version of the following article: Vega Moreno C, Miravalls\-Sierra E, Juli\'an Moreno G, L\'opez de Vergara JE, Maga\~na E, Aracil J. \textbf{On the design and performance evaluation of automatic traffic report generation systems with huge data volumes.} Int J Network Mgmt. 2018;e2044., which has been published in final form at \url{https://doi.org/10.1002/nem.2044}\\
This article may be used for non\-commercial purposes in accordance with Wiley Terms and Conditions for Use of Self\-Archived Versions.
}}}

\author[1,2]{Carlos Vega*}

\author[2]{Eduardo Miravalls-Sierra}

\author[1]{Guillermo Juli\'an-Moreno}

\author[1,2]{Jorge E. L\'opez de Vergara}

\author[1,3]{Eduardo Maga\~na}

\author[1,2]{Javier Aracil}

\authormark{Carlos Vega \textsc{et al}}

\address[1]{\orgdiv{NAUDIT}, \orgname{High Performance Computing and Networking, S.L.}, \orgaddress{\state{Parque Cient\'ifico de Madrid C/Faraday 7, 28049 Madrid}, \country{Spain}}\vspace{1em}}

\address[2]{\orgdiv{Departamento de Tecnolog\'ia Electr\'onica y de las Comunicaciones}, \orgname{Escuela Polit\'ecnica Superior, Universidad Aut\'onoma de Madrid}, \orgaddress{\state{C/Francisco Tom\'as y Valiente 11 (28049) Madrid}, \country{Spain}}\vspace{1em}}

\address[3]{\orgdiv{Departamento de Autom\'atica y Computaci\'on}, \orgname{Universidad P\'ublica de Navarra}, \orgaddress{\state{Campus Arrosadia, 31006, Pamplona, Navarra}, \country{Spain}}}

\corres{*Carlos Gonzalo Vega Moreno, \email{carlos.vega@naudit.es, carlosgonzalo.vega@predoc.uam.es}}

\presentaddress{Parque Cient\'ifico de Madrid. Calle Faraday, 7. 28049. Madrid}

\abstract[Summary]{In this paper we analyze the performance issues involved in the generation of automated traffic reports for large IT infrastructures. Such reports allows the IT manager to proactively detect possible abnormal situations and roll out the corresponding corrective actions. With the ever-increasing bandwidth of current networks, the design of automated traffic report generation systems is very challenging. In a first step, the  huge volumes of collected traffic are transformed into enriched flow records obtained from diverse collectors and dissectors. Then, such flow records, along with time series obtained from the raw traffic, are further processed to produce a usable report. As will be shown, the data volume in flow records is very large as well and requires careful selection of the Key Performance Indicators (KPIs) to be included in the report. 
In this regard, we discuss the use of high-level languages versus low-level approaches, in terms of speed and versatility. Furthermore, our design approach is targeted for rapid development in commodity hardware, which is essential to cost-effectively tackle demanding traffic analysis scenarios.}



\maketitle


\section{Introduction}\label{sec:introduction}

Service outages are one of the primary concerns of any datacenter or network manager. Using a sample of 69 datacenters in 43 institutions, it turned out that the average cost per minute of datacenter outage was 7,908 USD \cite{Ponemon2016}. To prevent such incidents, traffic analysis is a fundamental activity in datacenter or network management, which proactively helps identifying potential sources of trouble, before they happen.

Actually, the high-speed nature of current networks, together with the ever-increasing amount of new systems and applications, makes traffic analysis a difficult task. As it turns out, it is not a matter of analyzing a few Gbytes worth of traffic with a traffic sniffer any longer, but  digesting hundreds of GBytes of traffic and providing a quick diagnosis of the incident or proactive analysis. Such a digestion process primarily consists of funneling the incoming packets into flow records, which, in turn, serve to produce a usable report. \looseness=-1 

Such flow records do not only feature the usual netflow, IPFIX or sflow parameters, but also transport and application layer statistics. The former are mostly network and transport layer parameters related to volumetry, such as number of bytes transmitted between two IP endpoint through given TCP ports. The latter also include the HTTP or DNS response time or the number of TCP zero window announcements from the client or server, which have an impact in the perceived Quality of Service (QoS).

As the network bandwidth increases, so does the amount of flow records and the computing power required to process and summarize them  into a human readable report, thus becoming a very challenging task. In the past, the amount of flow records was manageable in a standard desktop and the analyst would simply use scripts to process the flow records, build the corresponding graphs and tables and finally produce the report. Nowadays, the amount of flow records has grown so large that automatic traffic reporting systems running on powerful servers are in order.

We believe that generation of traffic reports is cornerstone for traffic analysis. Needless to say, huge distributed databases, such as Hadoop systems, can be used for this purpose. However, such approach comes at a large hardware cost and is unpractical to deploy in a real network, taking into account that rack space and power are limited. Consequently, we advocate for a vertical scalability approach and use multi-core workstations instead of a fully distributed storage and processing system. Such multi-core workstation can also be used as a traffic capture probe with a high-speed driver~\cite{M3OMON}. Thus, the same workstation can be used for traffic capture, analysis and report generation purposes.Furthermore, we note that network operations' teams have strict budgetary restrictions and cannot afford a large processing infrastructure to produce traffic reports. Actually, the traffic report generation system should be fully optimized to make the most of commodity workstations.

Such automatic traffic report generation systems pose significant research challenges for their design and implementation, the most important of which is to deal with a huge volume of (enriched) flow records and produce a timely report in a cost-effective manner. Needless to say, reducing delivery time is a must for realistic network operations since, in case of an incident, a very fast response from the network operations center should follow.  

In this paper, we provide an extensive performance evaluation of the different processing and visualization techniques that can be used to automatically produce the report from the flow records. The state of the art shows considerable activity in packet capturing, and also in delivering flow records from the packets. However, as will be shown in  section~\ref{sec:art} there is scarce literature on the performance issues involved in the next step of transforming such flow records into usable information, namely, into a human readable report, especially when the volume is very large.

In this regard, we contribute to the state of the art by analyzing the design alternatives for automatic report generation from traffic flow records and provide an extensive trace-driven performance evaluation. To this end, we discuss the Key Performance Indicators (KPIs) involved in a traffic report and how to obtain them efficiently from the flow records. Traces have been obtained from real commercial environments, thus providing a realistic traffic load. Our findings serve as a guidance to other researchers and practitioners in the field in their efforts to implement automatic traffic report generation systems, capable to handle the vast amount of data seen in current networks and datacenters.

\subsection{State of the art} \label{sec:art}

Several approaches for traffic report generation systems have appeared in the literature, but to the best of our knowledge, either they are based on dashboards or they lack a performance evaluation with a large volume of traffic.

Those that are based on dashboards provide a first visual approximation of network activity, which is usually supplied in real-time. As such, a dashboard does not compare with a traffic report neither in traffic volume being treated nor in insights provided. Actually, dashboards are  interactive, providing search capabilities, zoom in and out features, etc., which demands that the data volume handled by the dashboard remains within reasonably small values. As a result, the aggregation level of the offered traffic statistics is necessarily coarse. In the realm of dashboard-based tools, works such as \textsf{ntop}~\cite{ntop} already provided at the beginning of the century a network traffic probe that captured packets to monitor network usage and displayed generic real-time statistics with a web-based interface. To this end, as noted before, packets were summarized into traffic flows (for instance at the transport level), which, in turn, serve to produce records that feed the dashboard.

Such an approach of transforming packets into flow records and then feeding the dashboard, possibly in real-time, is very challenging at high-speed. In this light, alternate approaches based on Netflow records are appealing because the data volume is drastically reduced, albeit only network-level statistics can be provided. For instance, NfSen project developed \textsf{NFDUMP} tools~\cite{nfdump} to capture and process these records, generating flow-based statistics that were also displayed in a web interface. In a similar vein, \textsf{SiLK}~\cite{silk} also provided a set of tools to collect, store and analyze network flows based on NetFlow or IPFIX. This type of records has been considered so useful that even a query language named NFQL~\cite{nfql} has been defined for them. However, we note that transport and application level statistics are missing in Netflow records. Consequently, they are targeted towards volumetric analysis, rather than quality of service assessment. \looseness=-1 

Given the limitations of dashboard-based approaches, especially those built on top of Netflow records, the research community focused on extending the Netflow capabilities by providing further insight into higher layers, such as \textsf{Tstat}~\cite{tstat}. This tool is able to generate records including TCP, UDP, and several upper layer protocols, such as TLS, RTP, DNS or HTTP. Moreover, it is a dashboard-based tool, namely it provides a web interface to visualize data. The authors of this tool have also studied how to aggregate statistical data from these records in~\cite{tstat_stats}. However, although they assess the error in their aggregation, they do not evaluate the time it takes to generate such statistics, especially with a large number of records. Furthermore, statistics are provided as tables and graphs, and not as a report.

We also note that there are commercial traffic sniffers (Netscout, Riverbed, Viavi, etc.) that provide traffic reports. However, such systems are not open-source and the internals are not described in the scientific literature.

Another example of tool that generates enriched records is \textsf{Tranalizer}~\cite{tranalizer}. Such tool can generate general statistics per protocol, and it can be used to feed standard tools such as RRD or Splunk, and even yield a PDF report, in the same way we present in our work. However, in their tests they have only documented the analysis of packet capture files as large as 44~GBytes, and reported the processing time for a small 1.3 GByte trace, whereas we are focusing on the analysis of very large captures with several TBytes worth of daily network traffic, as it usually happens in a real-world scenario. \looseness=-1 

\begin{figure*}[t!]
\centering
\includegraphics[width=\textwidth]{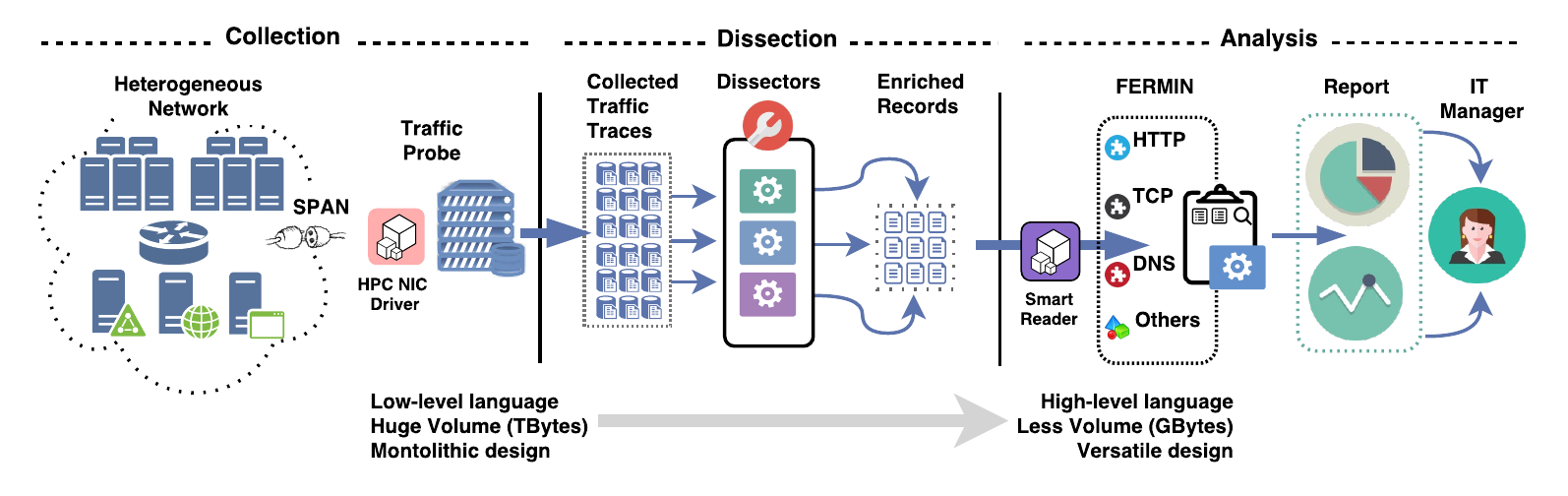}
\caption{Typical deployment scenario for traffic collection and analysis, depicting the different stages involved in the process.}
\label{fig:arch}
\vspace{-1em}
\end{figure*}

As shown, the state of the art features dashboard-based and reporting systems that handle a moderate volume of traffic traces or flow records. However, in a real-world datacenter, the collected data volume is much larger. For instance, in a large datacenter of a Spanish logistics company we collected a grand total of 18 TB worth of traffic (packets) and 298 GB worth of flow records on Thursday, November 9th, 2017. We stress that such traffic volumes are common in large datacenters~\cite{CISCO}. In our case, the datacenter management asked for a daily report that provided not only volumetric figures but also application layer statistics such as HTTP or DNS response time. Clearly, this is a different scale in terms of data volume, compared to the current SoA. In this paper, we analyze how to cost-effectively handle such a huge amount of data and produce a usable report. \looseness=-1 

In this paper, we provide insights into the design principles of automatic traffic report generation systems currently in production in large networks and datacenters~\cite{fermin_web}. The lessons learned help researchers and practitioners in the field to build efficient and useful automatic report generation systems.

\section{Methodology}
We consider a large IT infrastructure (datacenter, network, etc.) that performs IT service continuity management according to the ITIL methodology. To this end, a risk assessment (also known as risk analysis)   is performed to identify the assets, threats, vulnerabilities and countermeasures for each IT service. As a part of such an assessment, a traffic analysis is conducted on a daily or weekly basis, by means of an automated traffic analysis report. Then, the appropriate corrective actions are rolled out, and the alarm-based real-time monitoring systems are updated accordingly.

Figure~\ref{fig:arch} shows the different stages of automatic report generation. In the data collection stage, a traffic probe receives a copy of traffic from certain VLANs or ports of interest through a SPAN or port mirror. Then, several dissectors act on the captured traffic in order to deliver enriched flow records, which correspond to the different protocols in the traffic. For example, HTTP records that contain the client, server, URL, response time, response code, etc. Finally, the report is delivered using such enriched records as the data input. \looseness=-1 

The above stages can be performed in a single appliance or in several appliances, either in series or in parallel. We advocate for the use of a single appliance whenever possible. If several appliances are used, huge volumes of data must be moved from one appliance to the other. For example, if an appliance is used for data collection and dissection and a second appliance for analysis and report generation, respectively, all the enriched records must be transferred from the first appliance to the second one. Concerning parallel deployment of appliances, namely load balancing, we note that evenly distributing the incoming data is a challenging task. For example, in \cite{zipf} Shi \textit{et al}. the authors conclude that \textit{``under certain Zipf-like flow-size distributions, hashing alone is not able to balance workload''}, with regard to Internet flows.
Furthermore, cost increases if several appliances are used. Thus, we strive for the maximum efficiency in the data collection, dissection and analysis stages, to include all of them in the same host.

We note that the above three stages are not necessarily concurrent. A traffic trace can be collected during the day, and then, dissection and analysis can be performed at nighttime. For many use cases, automatic report generation systems are not aimed for real-time operation and alert triggering, but for long-term proactive analysis.

In what follows, we will provide insights into each of the aforementioned stages. In the next section, we describe how to perform traffic sniffing at multi-Gigabit-per-second rates and generate enriched records and aggregate statistics using high-speed dissectors. Then, the selection of Key Performance Indicators (KPIs) follows.

\subsection{Traffic capture and flow record generation}
Firstly, the corporate network traffic must be captured through either ad-hoc hardware or optimized drivers capable of receiving network traffic at 10 Gbit/s such as the HPCAP network driver~\cite{M3OMON} for Intel\textregistered~Ethernet NICs. As for traffic dissection, the traffic trace should be summarized into enriched records that contain amenable information for the analysis. In this regard, there is a trade-off between accuracy and real-time delivery of the enriched flow records.

We note that the on-line delivery of flow records is extremely challenging, if not impossible, at high-speed rates, especially if the application layer is to be inspected. Consequently, in some cases, the departure point for report generation is not the set of enriched records produced in real-time, but the traffic trace itself, which is subsequently processed to obtain enriched records. For example, a precise count of TCP retransmissions can only be obtained by fully reassembling the TCP connection and, then, searching for a possible retransmission for every incoming packet that belongs to the TCP connection. This is a very processing intensive task, which is unfeasible to perform live at high speed and should be performed offline.

We note that traffic analysis is an interactive top-down process. First, high-level statistics are analyzed and, then, the data is drilled-down in order to perform further investigation. As it turns out, chances are that the traffic trace has to be read {\em again}, which is costly in terms of I/O and adds significant processing time to the report generation. This is the case when inspecting the traffic from a particular IP that showed anomalous behavior in flow records, and calls for further analysis at the packet level.

In any case, the first step is traffic dissection and delivery of flow records, either online or offline. Traffic dissectors provide enriched records per service (HTTP, DNS, SQL, SMB, etc.), per protocol (TCP, UDP, ICMP), etc. Furthermore, they also provide time series that show the temporal evolution of a given statistic, for example, traffic in bits per second, in and out a given VLAN or subnetwork. As it turns out, there are high performance traffic dissectors available in the state or the art \cite{NDPI,HTTPD} and we will not focus on them in this article.

\subsection{Stateless and stateful flow records}

Generally speaking, enriched flow records can be classified into stateless and stateful.

\subsubsection{Stateless flow records}
These flow records usually feature absolute counters that are not context-aware, namely the estimation of which does not require state. Typically, they report global statistics such as the number of packets or bytes transmitted per flow. Usually, such metrics can be delivered online, even at high rates. \looseness=-1 

\subsubsection{Stateful flow records}
In addition, more elaborated metrics require awareness of previous states, such as those related to conversations, connections or sessions. Such metrics require higher amount of resources and are more time-consuming, hampering real-time delivery. In the worst-case scenario, they may require multiple read passes over the analyzed trace.

For example, some traffic probes\cite{M3OMON} produce enriched records in real-time that provide the number of zero-window announcements from a given client or server in a TCP flow. However, we note that not only the zero-window announcement events matter, but the maximum time that the client or server was not opening that window. If a given client or server announces a zero-window just once in a TCP connection, then it does not matter that much. Conversely, if such an announcement is held for 10 seconds, chances are that the server is heavily saturated.
However, the client or server blocked time (i.e. time it takes to the host to open the window back again) should be obtained off-line, to reduce the probe processing requirements while capturing and storing traffic at high speed.

\section{Data volumes}
\label{sec:design}

\begin{figure*}[b]
\vspace{-2em}
  \begin{subfigure}[tbp]{0.60\textwidth}
\captionsetup{font=normalsize={bf,sf}}
\captionsetup[sub]{font=small={bf,sf}}
    \centering
\includegraphics[width=\columnwidth]{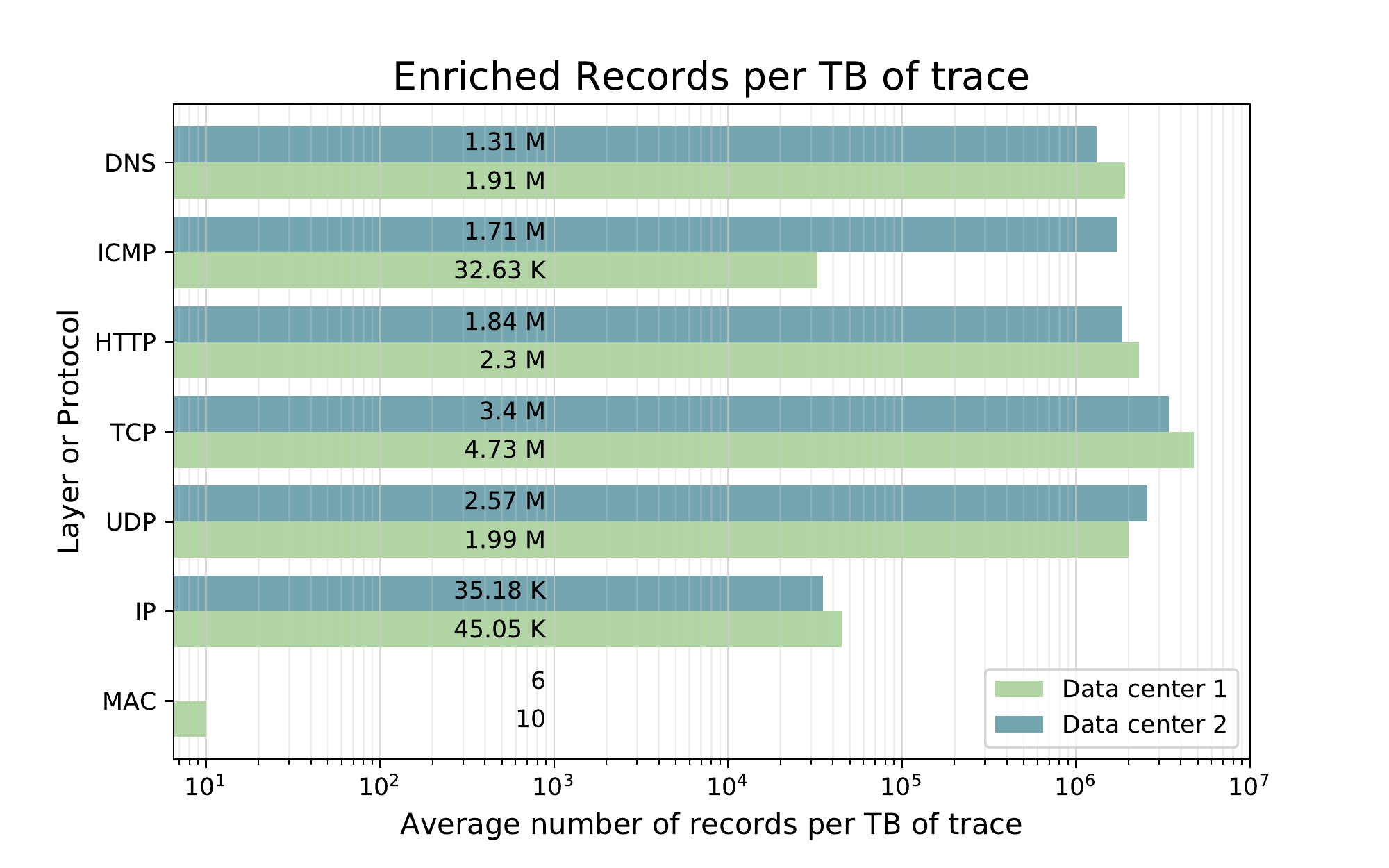}
\caption{Enriched records per TB of trace from different companies. \\ Labels indicate the value of each bar.}
\label{fig:perTB}
  \end{subfigure}
  \hfill 
  \begin{subfigure}[tbp]{0.42\textwidth}
  \captionsetup{font=normalsize={bf,sf}}
\captionsetup[sub]{font=small={bf,sf}}
      \centering
   \includegraphics[width=\columnwidth]{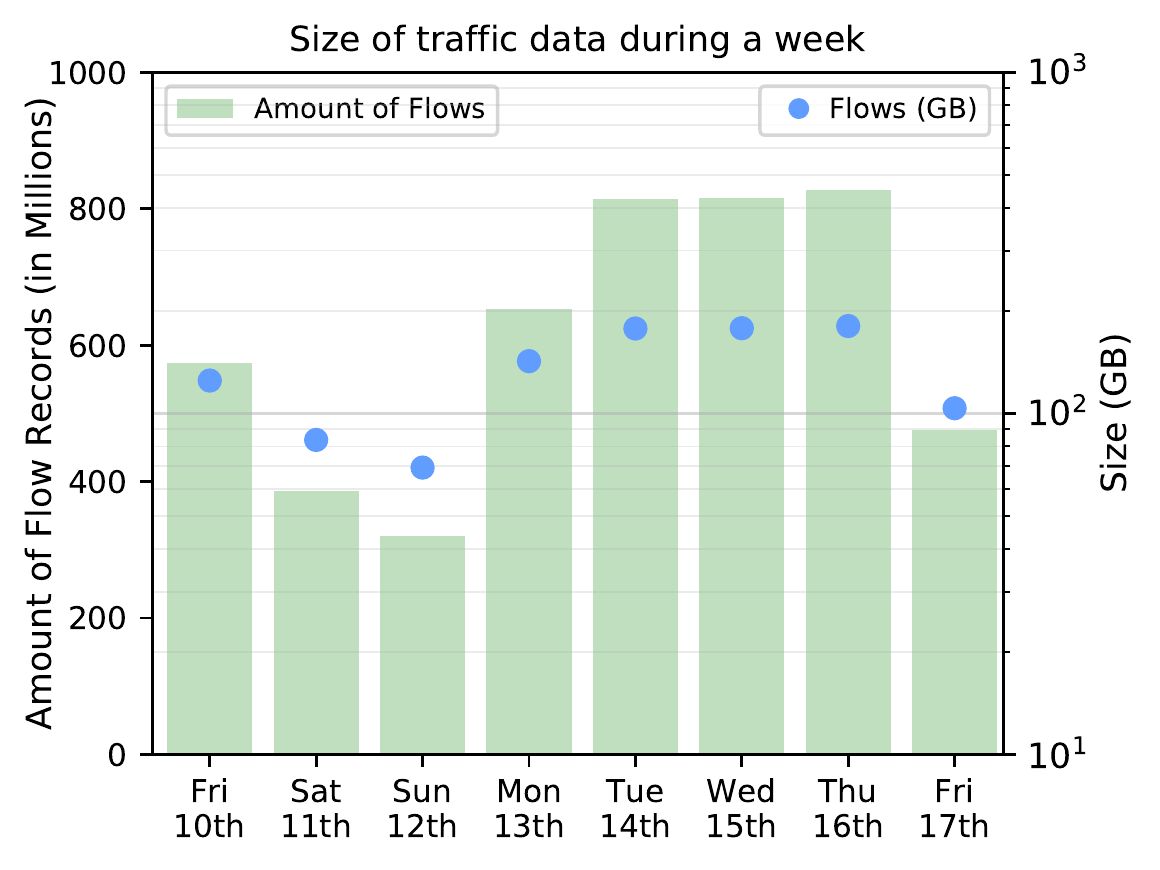}
    \caption{Amount of flows records for each weekday and their corresponding size in GigaBytes. Data was collected in a large datacenter of a Spanish logistics company.}
    \label{fig:raws_flows_week}
  \end{subfigure}
  \caption{Statistics from different enterprise networks.}
\end{figure*}

Figure~\ref{fig:perTB} shows the volume of enriched records per TB of traffic trace for two different datacenters, for MAC/IP/TCP/UDP/HTTP flows with an average record length of $29\pm44$ fields and $195\pm121$ bytes per record. The length in bytes of the enriched records spans from less than 10 for MAC to more than 1000 for TCP. The former usually consists of a few counters reporting the number of packets or bytes transmitted from each endpoint, while the latter has dozens of metrics for both directions, such as number of retransmissions or duplicates. 
Consequently, a careful capacity planning of the host memory is in order. To this end, knowledge of the approximate volume of enriched records per TB of traffic trace becomes fundamental. Figure \ref{fig:raws_flows_week} shows the amount of flow records (similar to ~\cite[pg. 10]{M3OMON} flows) collected per day in a large Spanish logistics company datacenter, showing that a number of 600-800 millions of flow records per day can be expected, which make up 200 GBytes approximately in size.

The figure shows that for a large datacenter several tens of millions of records can be produced per day. Thus, a trade-off between high-level programmability and efficiency has to be sought.

\begin{figure*}[b]
  \begin{subfigure}[tbp]{0.49\textwidth}
\captionsetup{font=normalsize={bf,sf}}
\captionsetup[sub]{font=small={bf,sf}}
    \centering
    \includegraphics[width=\columnwidth]{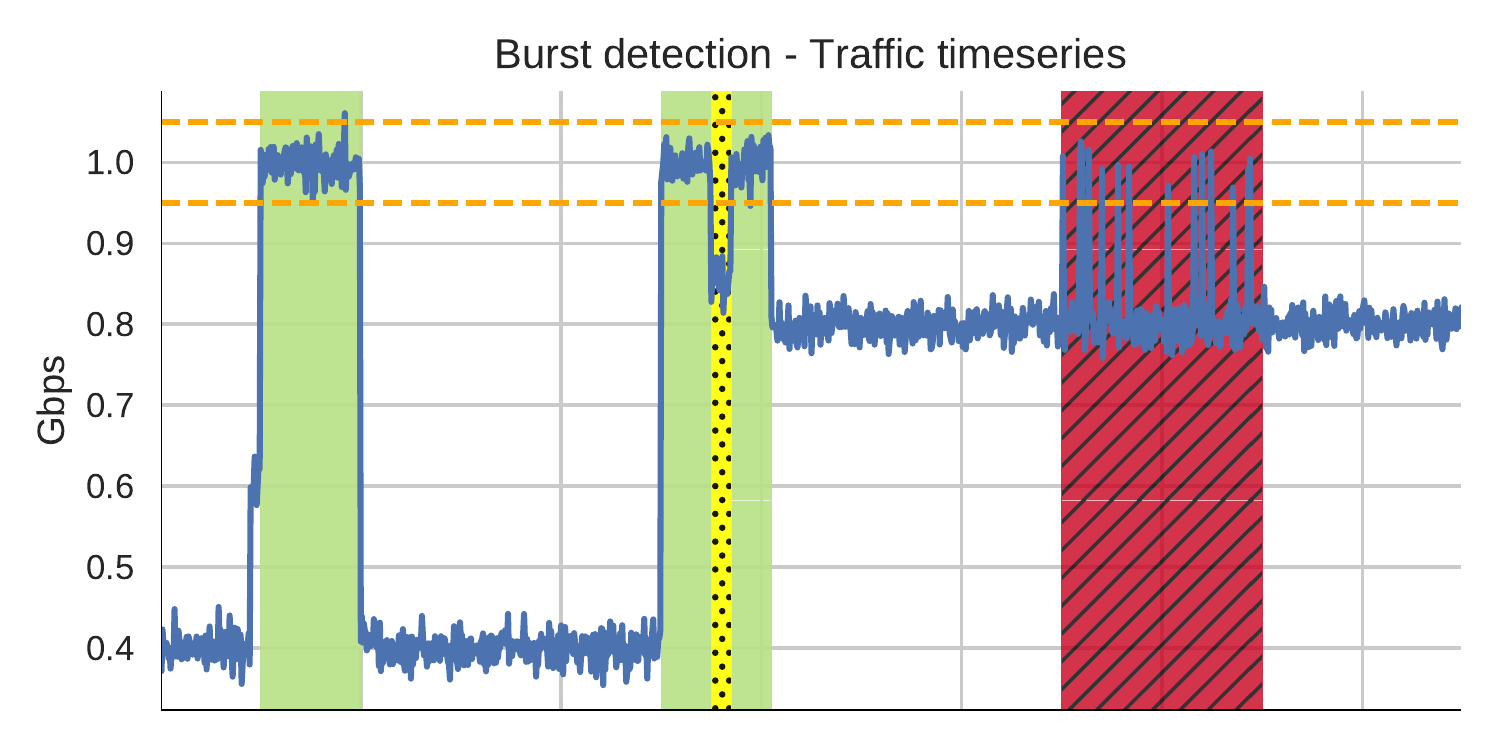}
    \caption{Example situations for the burst detection algorithm: detected bursts (green), extended bursts (dotted yellow) and rejected bursts because of low duration and/or average speed (dashed red). Orange lines represent the detection thresholds.}
    \label{fig:Bursts}
  \end{subfigure}
  \hfill 
  \begin{subfigure}[tbp]{0.49\textwidth}
  \captionsetup{font=normalsize={bf,sf}}
\captionsetup[sub]{font=small={bf,sf}}
      \centering
      \includegraphics[width=\columnwidth,height=0.4\columnwidth]{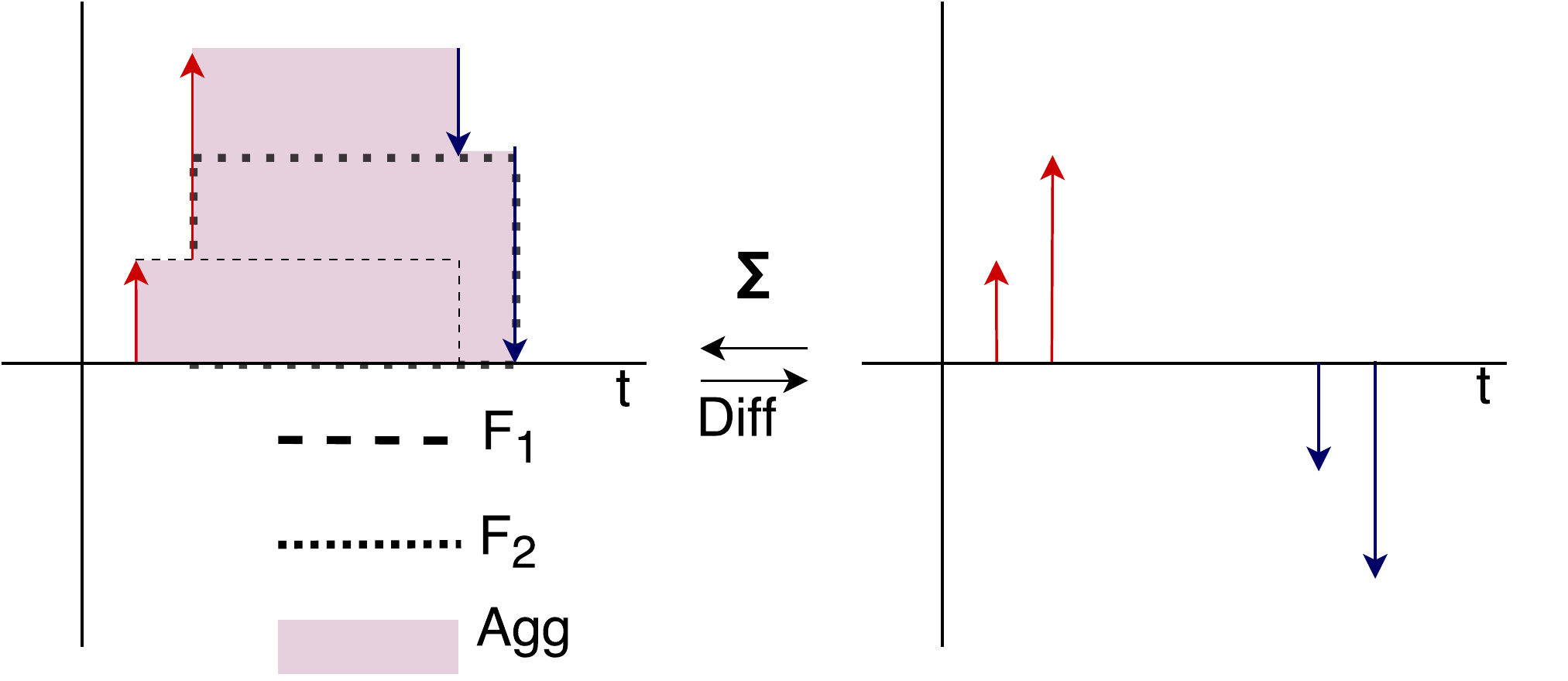}
      \vspace{0.2cm}
      \caption{Reconstruction of a time series. The figure depicts an example with two flows, in which the reconstruction is performed by using the cumulative sum of these impulses.}
      \label{fig:cumsum}
  \end{subfigure}
  \caption{Representation of some methods used for the analysis of time series.}
  \label{fig:algorithms}
\end{figure*}

Processing speed is a crucial requirement for automatic report generation. Implementation-wise, the most important issues in such automatic report generation are single-pass analysis and arrangement of the data files in memory, which, in turn, impose restrictions to the KPIs to be included in the report.Single-pass analysis entails that the traffic trace is read from hard disk just once, which gets dissected into records as it is read, and, then, the necessary graphs and tables for the report are produced. We note that reading TBytes of data from hard disk is very time consuming. Consequently, the less read passes over the trace the better performance. Thus, most of the KPIs should be obtained from enriched records. For example, obtaining a detailed time series for a given host, which has been identified in the enriched records analysis, \textbf{implies reading the traffic trace a second time}, which is very costly in terms of time.

Concerning information arrangement, the task comprises extracting the fields of interest from the enriched records and keeping them in memory. As noted before, in a large datacenter, a one-day worth of traffic may be around several TBytes and the enriched records may occupy several tens of GBytes. In order to obtain the report graphs and tables, such enriched records are heavily accessed during the report generation process. Therefore, they should be cached in memory and not retrieved from disk if possible, to avoid penalizing performance significantly.

\section{Main Key Performance Indicators}

As noted before, the selection of KPIs is conditioned by the limiting factors stated above of single-pass analysis and arrangement of data files in memory. Needless to say, the basic volumetry indicators (top clients and servers, conversations, ports, etc.) can be easily obtained from the enriched records using hash tables stored in memory. More elaborate KPIs may actually require several passes to the enriched records, which could penalize performance.

We will not provide a description of all the KPIs for the sake of brevity. Instead, we will focus on original KPIs that can be obtained with moderate processing resources, and that provide valuable information to an IT manager.

\subsection{Burst analysis and identification of root cause}

Burst are useful to the IT manager to identify potential saturations, for example, a saturated link shows a sustained traffic burst at its line rate. However, time resolution is cornerstone in order to identify short bursts and perform the appropriate proactive actions before they become significant. \looseness=-1 

In order to identify such bursts (see Fig.~\ref{fig:Bursts}) or plateaus, we use the time series of bits per second of the whole trace, with one-second time resolution, which is obtained from the first pass to the trace. We note that such a time series cannot be obtained from the flow records. Then, a simple heuristic for burst detection based on throughput threshold and duration above threshold is adopted. Finally, we provide the candidate burst root cause, which is of interest to identify what happened.

To this end, we perform the matching of a set of time series from different metrics (HTTP response time, number of UDP flows, etc.) against each of the bursts from the bits per second time series of the trace, looking for correlations in the variation of the metrics. The metric with highest variability in the burst time interval is chosen as the candidate root cause (as in principal component analysis), and the endpoints identified in the root cause metrics are highlighted in the report. This variability is computed using a rolling centered window, assigning to each window the standard deviation of its values normalized by its mean.

\begin{figure*}[b]
\vspace{-1em}
  \begin{subfigure}[tbp]{0.45\textwidth}
\captionsetup{font=normalsize={bf,sf}}
\captionsetup[sub]{font=small={bf,sf}}
    \centering
\includegraphics[width=\columnwidth]{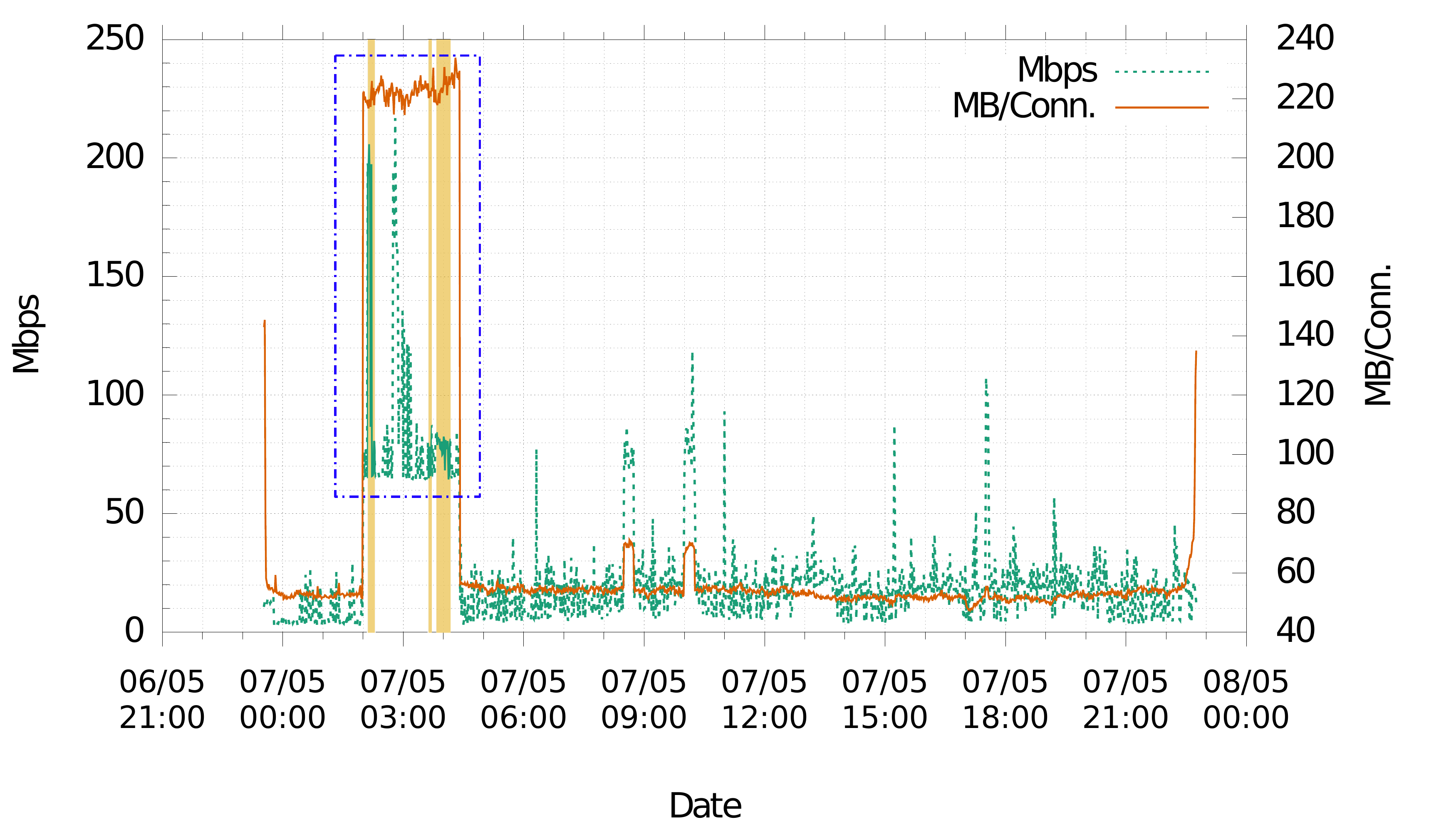}
    \caption{In the span marked with a dotted blue rectangle, \\ a traffic burst (dotted green) is caused by a sudden \\ increase of the traffic per connection (orange line).}
    \label{fig:mbperconn}
  \end{subfigure}
  \hfill 
  \begin{subfigure}[tbp]{0.55\textwidth}
  \captionsetup{font=normalsize={bf,sf}}
\captionsetup[sub]{font=small={bf,sf}}
      \centering
   \includegraphics[width=\columnwidth]{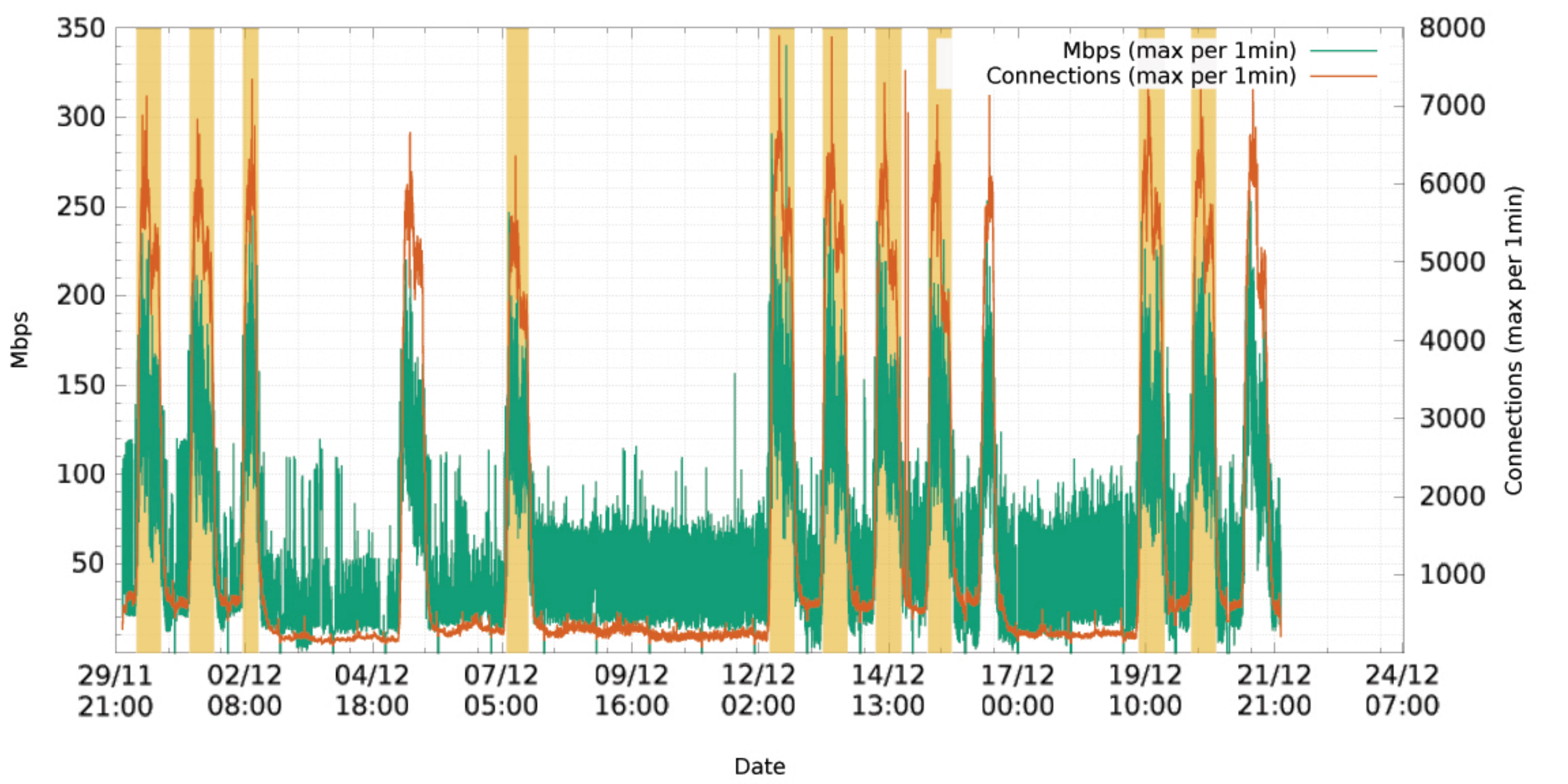}
    \caption{Burst detection in a real report and corresponding cause time series. Highlighted regions correspond to bursts detected with a rate of 100~Mbps.}
        \label{fig:BurstsFerr}
  \end{subfigure}
  \caption{Real examples of traffic bursts.}
\end{figure*}

For instance, Figure~\ref{fig:mbperconn} depicts an example of traffic burst. Bursts are marked with a yellow area, and in this case, the candidate root cause of the traffic burst is an increase of traffic per connection (MB/Connection line). As shown, connections unexpectedly grow in size, which may be indicative of an anomaly.

Figure~\ref{fig:BurstsFerr} shows another example in which  several bursts are caused by an increased number of connections. Based on these correlations, the software isolates the flows that were observed during the burst time and shows the clients with most connections in that period. This allows the manager to see not only why a burst happened, but also which nodes in the network were responsible for it.

\subsection{Time series reshaping from enriched records}

Time series are essential to understand the dynamics of a given statistics, such as for example evolution of consumed bandwidth. Nevertheless, we note that an accurate time series cannot be obtained from enriched records~\cite{NetFlowLossWin2002}. For example, if we wish to obtain the traffic in bits per second in and out from a given IP and port, the enriched records will only provide the average traffic per flow, for those flows that have that IP and port as an endpoint.

Therefore, we can recreate the time series we are looking for by adding up such averages as step functions in the intervals where the flow is alive (see Fig.~\ref{fig:cumsum}). This method assumes that the bandwidth consumed during the flow lifetime is uniform.  Then, we add up the step functions to yield an approximate time series for the specific metric. However, this is just an approximation of the real time series of the input and output traffic for a given set of IP addresses and port numbers.
In fact, the resulting traffic from the composition of such step functions is less \textit{bursty} than the real one. Hence, we cannot display the traffic peaks accurately, which is cornerstone for some services. For example, detection of peaks in Citrix~\cite{CITRIX} traffic is a very important issue as such peaks produce instantaneous queuing delays that are noticeable by users, even in small time scales.

In this example, we note that the time series to be obtained cannot be programmed beforehand. First, a flow-based analysis is performed that provides the hosts with potential problems (for instance a high retransmission rate per flow). Then, a detailed traffic time series can be obtained for that particular host in order to look for possible traffic anomalies, such as micro-outages in the scale of seconds, that cannot be derived from the enriched records directly. However, the latter requires a second pass to the trace.  \looseness=-1 

\begin{table}[t]
\setstretch{1}\fontsize{9}{11}\selectfont
\centering
\caption{Key Performance Indicators used for the RAG tables and their corresponding conditions. When the conditions are fulfilled, a value is added up to the server score. Servers with highest scores are shown first in the RAG table.}
\label{KPItable}
\def\arraystretch{1.2}
\begin{tabular}{|C{1cm}|C{3.8cm}|C{7.5cm}|C{3.5cm}|}
\hline
\textbf{Protocol}        & \textbf{KPI}                     & \textbf{Trigger value}                                                                                                     & \textbf{Score to be added}                          \\ \hline
\multirow{5}{*}{\begin{tabular}[c]{@{}c@{}}\textbf{TCP}\\ \textbf{servers}\end{tabular}}              & Duplicate ACK Src. To Dst.       & 5\% per connection                                                                                                     & 2 per unit                                    \\ \cline{2-4}
                                                                                     & Duplicate ACK Dst. To Src.       & 5\% per connection                                                                                                     & 2 per unit                                    \\ \cline{2-4}
                                                                                     & Retransmissions Src. To Dst.     & 5\% per connection                                                                                                     & 2 per unit                                    \\ \cline{2-4}
                                                                                     & Retransmissions Dst. To Src.     & 5\% per connection                                                                                                     & 2 per unit                                    \\ \cline{2-4}
                                                                                     & Zero Window Dst. To Src.         & 5\% per connection                                                                                                     & 2 per unit                                    \\ \cline{2-4}
                                                                                     & Downtimes                        & SYNs from source to server without packets from the server and no data packets from the source during 90\% of the time & 25 if true                                    \\ \cline{2-4}
                                                                                     & Connection Establishment Time (s) & CET $\geq$ 0.1s during 5 consecutive minutes OR a 5 min spike 10 times higher than the mean value                           & 50 if true                                    \\ \cline{2-4}
                                                                                     & Round-Trip Time (s)              & RTT $\geq$ 1s  during 5 consecutive minutes OR a 5 min spike 10 times higher than the mean value                            & 50 if true                                    \\ \cline{2-4}
                                                                                     & Ignored / Denied SYNs            & Used to measure the importance of the server                                                                           & 0.1 per unit                                  \\ \cline{2-4}
                                                                                     & Number of Connections            & Used to measure the importance of the server-1 if there are no records with SYNACK but more than 1000 with SYN         & 0.01 per unit OR a total of 10 if -1          \\ \cline{2-4}
                                                                                     & Bytes transmitted                & Used to measure the importance of the IP server                                                                        & 0.1 per unit                                  \\ \hline
\multirow{6}{*}[-1em]{\begin{tabular}[c]{@{}c@{}}\textbf{HTTP}\\ \textbf{servers} \\ \textbf{and} \\ \textbf{ports}\end{tabular}} & Server Errors (\%)               & 5\% of Server Errors relative to the server                                                                            & \multirow{2}{*}{\begin{tabular}[c]{@{}c@{}} $\frac{3\; \cdot \; Server\; Errors\; \left( \% \right)\; \cdot Transactions\; }{100}$ \\ \end{tabular}} \\[3ex] \cline{2-4}
                                                                                     & Client Errors (\%)               & 20\% of Client Errors relative to the server                                                                           & \multirow{2}{*}{\begin{tabular}[c]{@{}c@{}} $\frac{Client\; Errors\; \left( \% \right)\; \cdot Transactions\; }{100}$ \\ \end{tabular}} \\[3ex] \cline{2-4}
                                                                                     & Median Response Time (s)         & Greater than or equal to 0.1 seconds                                                                                   & 50 if true                                    \\ \cline{2-4}
                                                                                     & Avg. Response Time (s)        & Greater than or equal to 0.5 seconds                                                                                   & 50 if true                                    \\ \cline{2-4}
                                                                                     & Acc. Response Time (\%)   & Used to measure the importance of the HTTP server and its corresponding port                                           & 2 per unit                                    \\ \cline{2-4}
                                                                                     & Transactions                     & Used to measure the importance of the HTTP server and its corresponding port                                           & 1 per unit                                    \\ \hline
\multirow{5}{*}{\begin{tabular}[c]{@{}c@{}}\textbf{DNS}\\ \textbf{servers}\end{tabular}}               & Errors (\%)                      & 5\% of Errors relative to the server                                                                                   & 2 per unit                                    \\ \cline{2-4}
                                                                                     & Median Response Time (ms)        & Greater than or equal to  100 ms                                                                                       & 50 if true                                    \\ \cline{2-4}
                                                                                     & Avg. Response Time (ms)          & Greater than or equal to 500 ms                                                                                        & 50 if true                                    \\ \cline{2-4}
                                                                                     & Acc. Time (\%)                   & Used to measure the importance of the DNS server                                                                       & 1 per unit                                    \\ \cline{2-4}
                                                                                     & Transactions                     & Used to measure the importance of the DNS server                                                                       & 1 per unit                                    \\ \hline
\end{tabular}
\end{table}

\subsection{Red-Amber-Green summaries}

A Red-Amber-Green (RAG) diagram provides a high-level performance evaluation of a given system, service or link. To this end, summaries of specific metrics are conducted for different layers and protocols such as DNS, TCP, HTTP, etc. For example, the following TCP metrics are taken into account for a TCP service RAG analysis: zero-window announcements, retransmissions, duplicate ACKs, SYN attempts, downtimes, RTT, connection establishment time, and the number of connections and the relative contribution in bytes to the TCP traffic.

Several thresholds and heuristics are used to characterize the server health in different groups, for instance, a time interval between the SYN-ACK and the ACK in the TCP handshakes serves to estimate the RTT. If such RTT is higher than 1 second for 5 consecutive minutes or shows a significant departure from the mean (e.g. a sudden tenfold RTT increase), then an anomaly is signaled for the RTT metric. 

Table~\ref{KPItable} summarizes most of the KPIs used for the RAG tables of the traffic report, together with a brief description of the trigger values tthat indicate an anomalous behavior of the KPI metric. Each of these trigger values, if fulfilled, adds up a score value to the corresponding server. The final RAG table features the servers sorted by the described score and colored according to the severity of their status.

\section{Implementation issues}

Clearly, the collection, dissection and analysis stages have very different needs from the programming perspective. As we climb from the wire to the charts (see Fig,.~\ref{fig:arch}) we tend to aggregate the information, from the individual packets to the enriched records and finally the charts in the automatic report. During the collection process, kernel-level ad-hoc software provides maximum performance to avoid packet loss during the capture process. Afterwards, high-speed dissectors swiftly extract the information, possibly offline. \looseness=-1 

These two first stages usually prioritize non-functional requirements such as performance, resilience, efficiency and stability and tend to evolve slowly with few changes during its lifecycle due to its monolithic and fit-for-purpose design. For example, protocol traffic dissectors are not modified often since protocols (e.g. TCP) are neither changed frequently.

However, the analysis process requires higher level of abstraction and complex functions more prone to blow with the wind depending on the analysts and IT manager needs. These are often functional requirements such as analyzing new metrics or drawing additional charts. Thus, such software layers usually require non-functional requirements such as extensibility, modularity and versatility.

\begin{figure}[b!]
\centering
\includegraphics[width=0.75\columnwidth]{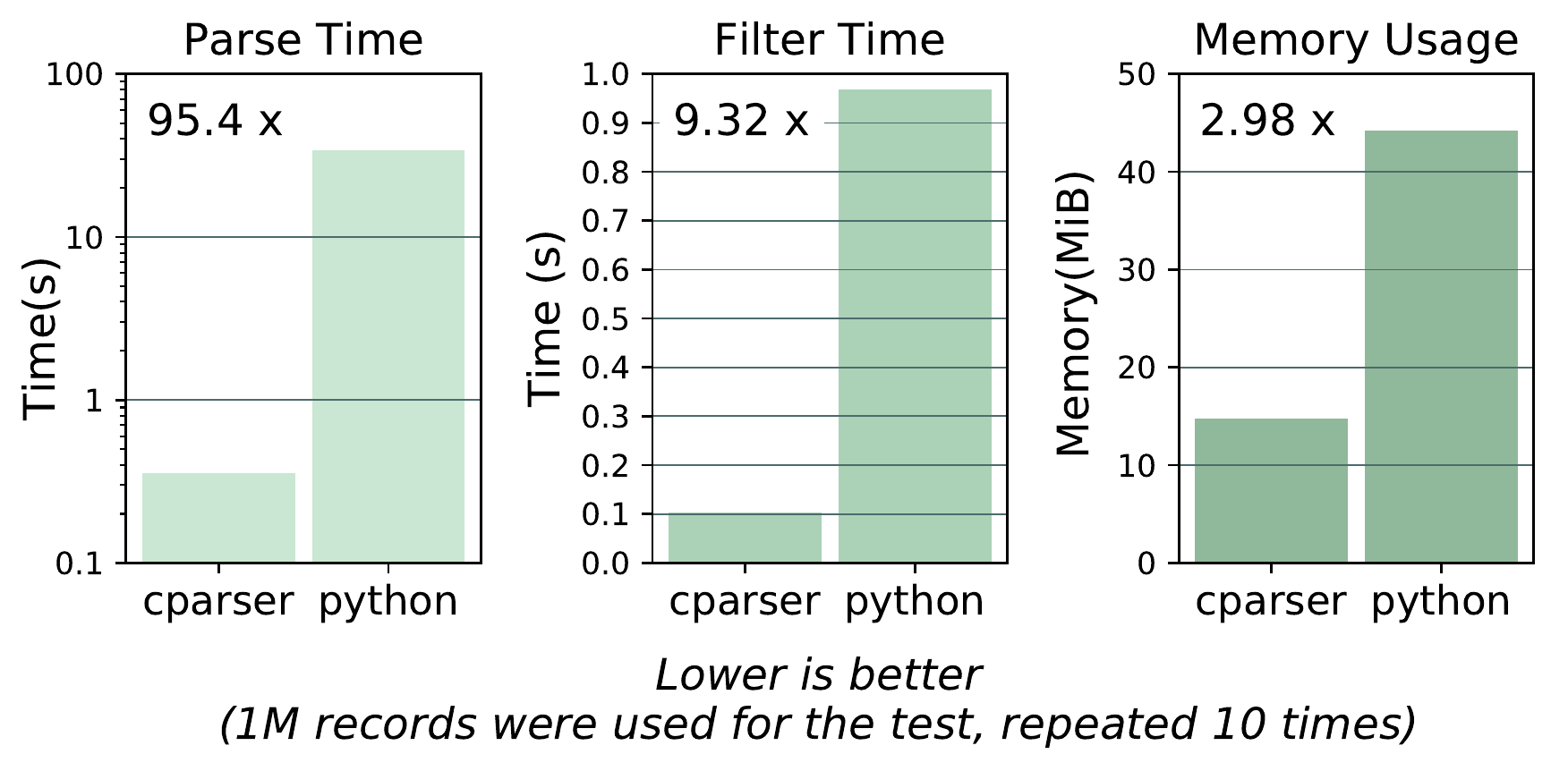}
\captionsetup{justification=centering}
\caption{Performance comparison between \textsf{cparser} and standard Python I/O libraries.}
\label{fig:cparser}
\vspace{2em}
\end{figure}

\subsection{Analysis of the enriched records}

Hence, in order to tackle with the various analysis requirements mentioned above, we have considered \emph{Python} due to its versatility and portability, making use of high-level statistical libraries such as \emph{Pandas}~\cite{pandas_web} and \emph{Numpy}~\cite{numpy_web}, which allowed for a rapid development of our system while keeping good performance. 

Alternate programming language options are C\# or Java, but they lack the flexibility to implement new metrics as required.  Clearly, compiled languages such as C or C++ would definitely provide better performance but at the cost of excessive development time. \looseness=-1 

However, we can avoid that excessive development time and take advantage of the extra performance and leaner footprint of the C language by using the C-Python interface in performance-critical sections. This has proven extremely useful in our solution, as we found that Python had an unacceptable poor performance (both in terms of memory usage and speed) when reading all the records from the files, and that alternatives within Pandas or Numpy could not support all our data types.

To circumvent the above performance barrier, we developed a C module (\textsf{cparser}~\cite{cparser_web}) that provided faster reading of the records and lower memory usage compared with Python (see Fig.~\ref{fig:cparser}). On top of this module, we implemented another module that provides common data operations (filters, aggregations, maps, top values, etc) that may be chained efficiently, aching the data and deferring read operations until needed. The \textsf{cparser} code is available to the community, as an additional contribution of the paper.\looseness=-1 

Figure~\ref{fig:cparser} shows the performance comparison between \textsf{cparser} and the standard Python I/O libraries. 

The figure shows the read times for one million records, with 10 runs per experiment. As shown, the decrease in the read time is striking with our \textsf{cparser} solution. \looseness=-1 

The improved performance brought by \textsf{cparser} implies that the time spent during the record parsing is approximately 1\% of the total execution time (8 minutes). Without \textsf{cparser}, our heaviest reports  took a full day to generate. Instead,  \textsf{cparser} reduces the generation time to 9 hours and 30 minutes.

\subsubsection{Visualizations and report generation}

In order to deliver useful visualizations it is always good practice to follow simple rules~\cite{10RULES} about the appropriate way to represent data. Since many different kind of charts and visualization libraries are available, it is not a trivial task to choose the right way to display information, especially when we require chart automation. We have studied and extensively tested the features of the most popular visualization libraries such as \emph{bokeh}~\cite{bokeh_web}, \emph{ggplot}~\cite{ggplot_web} (based on \textit{The Grammar of Graphics}\cite{ggrammar}), \emph{gnuplot}\cite{gnuplot_web}, \emph{tikz}~\cite{tikz_web}, \emph{matplotlib}\cite{matplotlib_web} or \emph{seaborn}\cite{seaborn_web} (based on matplotlib). We opted for \emph{gnuplot} and \emph{seaborn} for the main charts and \emph{tikz} for the topology graphs. 

The latter are well suited for our requirements of speed, integration, versatility, and customization. We were able to produce a large variety of custom charts such as survival distribution function plots, boxplots, violinplots, time series, pie charts and topology graphs that reshaped themselves depending on the data. For example, they allow resampling any time series in a given time span.

The final step in producing the report can be performed by means of a typesetting language such as \LaTeX, which is open source and allows creating a formatted document from a text file. Thanks to our modular design, we do not disregard other output formats such as plain text in console that could link the automatic traffic generation system with other management tools.

\subsubsection{Future of high-level languages in automatic reporting}

We look forward to the progress of projects such as PyPy\cite{pypy_web}, Numba\cite{numba_web} or Grumpy~\cite{grumpy_web} which aim to dilute the boundaries between high and low level programming languages, bringing together the strong abstraction of the former while keeping the high performance of the latter. We note that they still require further refinement and wider compatibility but the trend of high level languages with faster just-in-time compilers looks promising for the upcoming challenges of automatic traffic reporting and network management.

\begin{figure}[b]
\centering
\includegraphics[width=0.75\columnwidth]{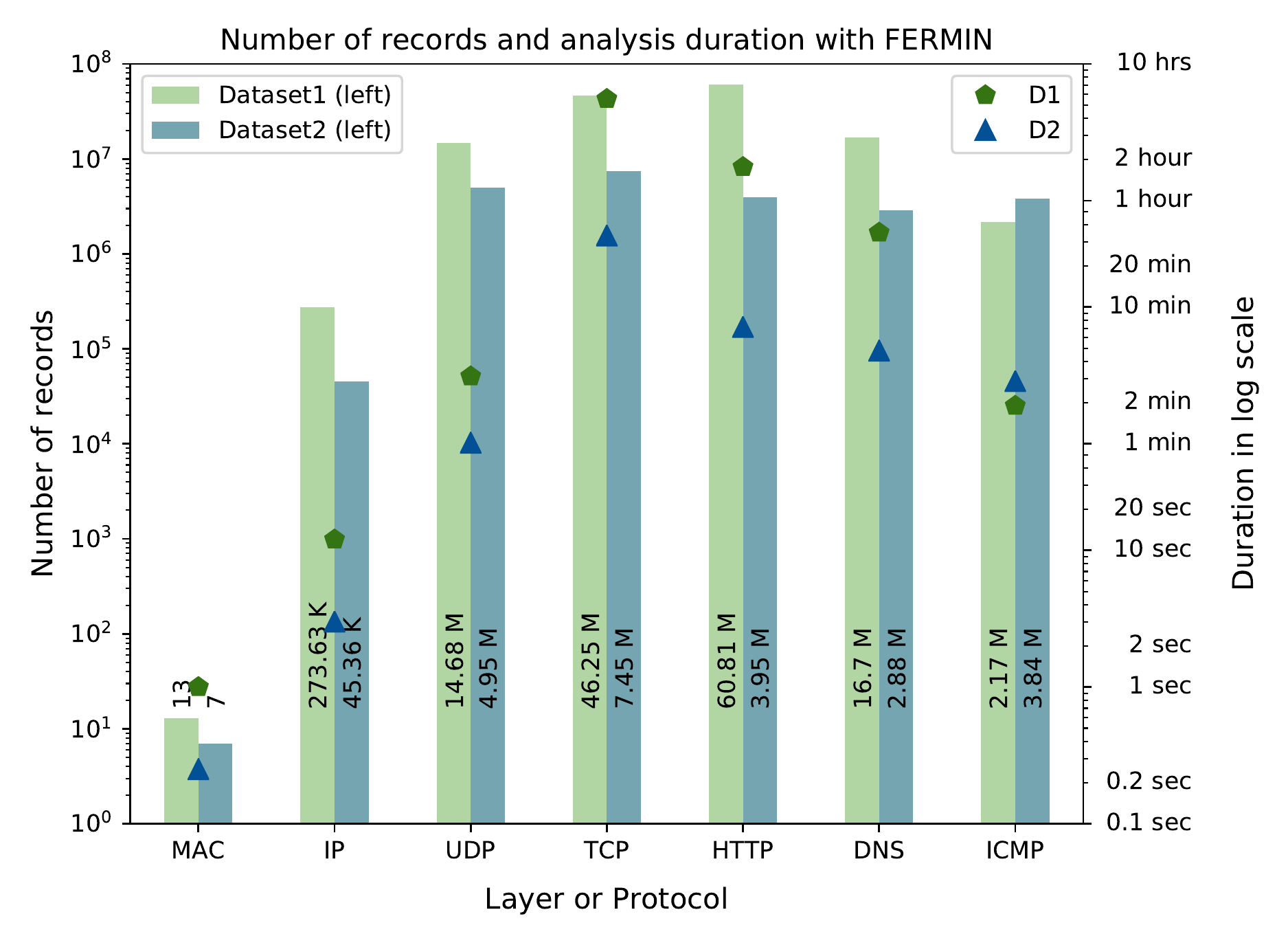}
\caption{Amount of enriched records and their corresponding processing time for each layer/protocol in different datasets.}
\label{fig:stage_times}
\end{figure}

\section{Evaluation}
\label{sec:meth}

For performance evaluation purposes, we collected two groups of real network traces from different multinational corporations.  

These two groups have different traffic characteristics. The 6.9 TB (10,036 Million Packets) of traffic traces from \textbf{Dataset~1} were captured from two different interfaces at the edge of a company network. \textbf{Dataset~2} was captured in the core distribution layer of a different company network, providing 2.1 TB (3,693 Million Packets) of traffic traces. \looseness=-1 

\begin{figure*}[b]
\centering
\begin{subfigure}{0.5\columnwidth}
  \captionsetup{font=normalsize={bf,sf}}
\captionsetup[sub]{font=small={bf,sf}}
 \centering
 \includegraphics[width=\columnwidth]{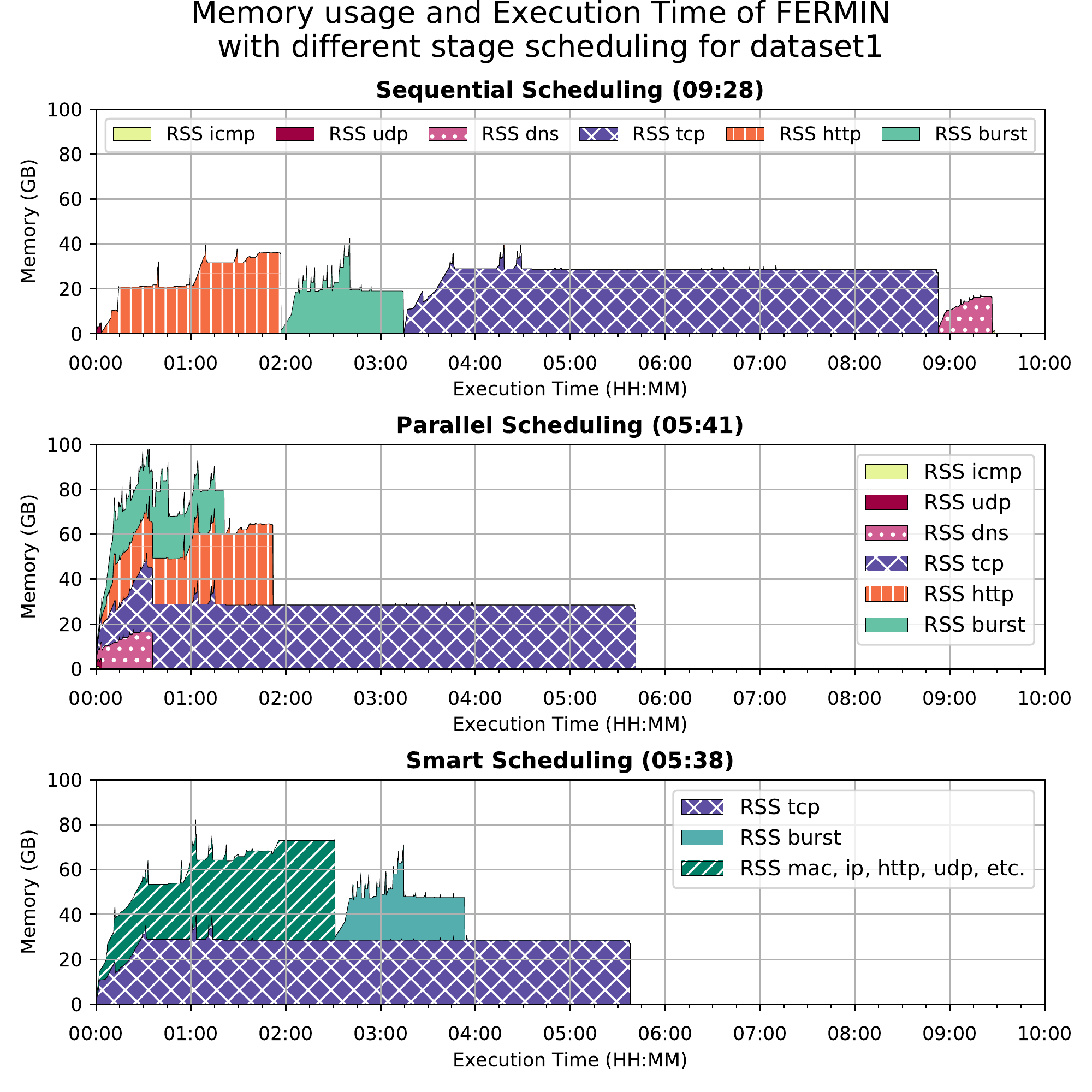}
 \caption{Dataset 1}   \end{subfigure}%
~
\begin{subfigure}{0.5\columnwidth}
  \captionsetup{font=normalsize={bf,sf}}
\captionsetup[sub]{font=small={bf,sf}}
 \centering
 \includegraphics[width=\columnwidth]{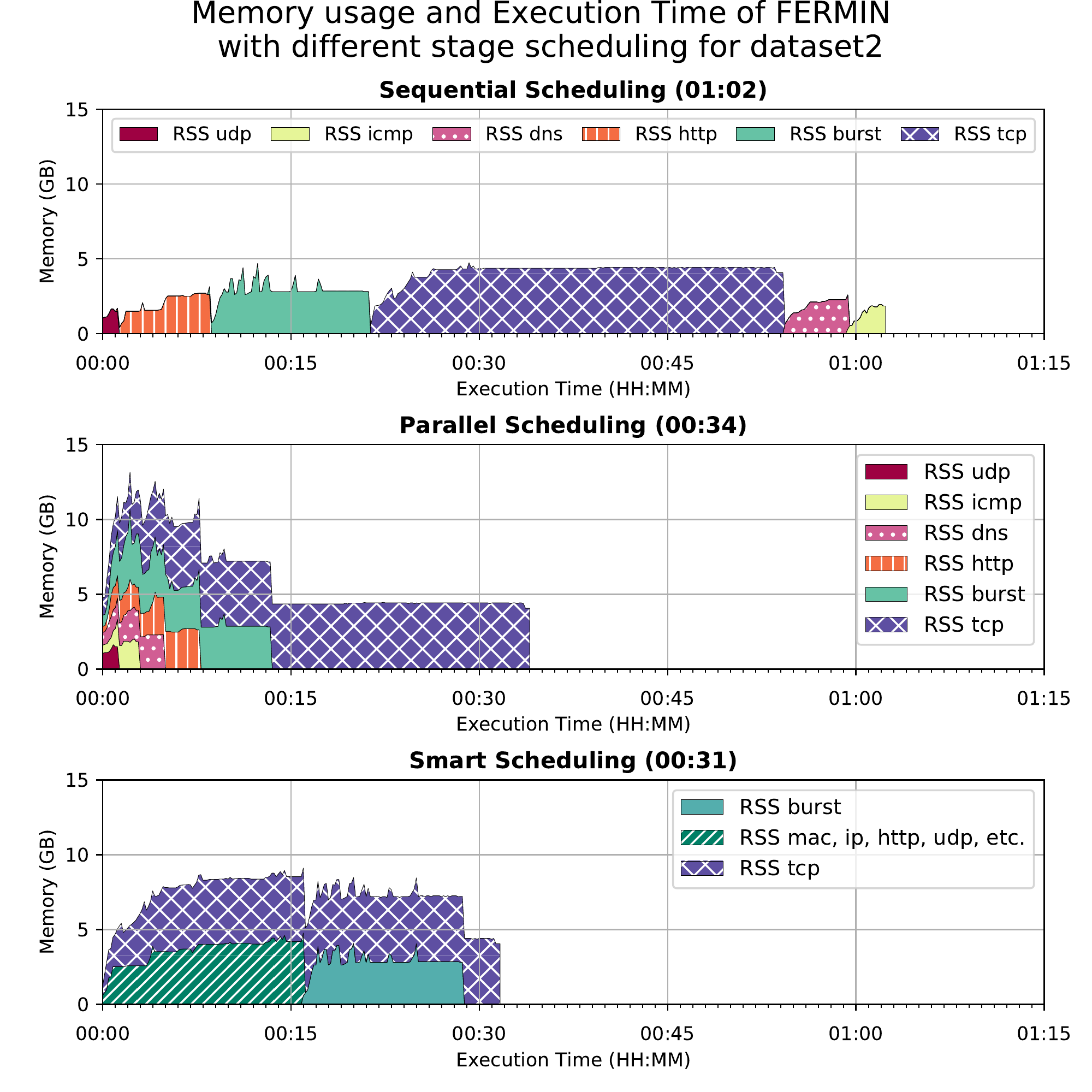}
 \caption{Dataset 2}   \end{subfigure}%
\vspace{-0.5em}
\caption{Memory usage and execution time for different stage schedulings. Short stages are not shown.}
\label{fig:schedulings}
\end{figure*}

\subsection{Dataset preparation}

Following the collection of the network traces, they were summarized into enriched records\footnote{Examples available on request.} by specialized dissectors. Such records contain different statistics depending on the dissected layer, and the amount of information is very variable as well. The MAC or IP conversations associated records naturally provide coarser information than the TCP counterpart. Application protocols usually have more text-based fields such as the HTTP URI or the requested host during a DNS query, with an impact on both the record length and memory consumption during the analysis.

The resulting enriched records are also depicted in Figure~\ref{fig:stage_times}, showing the amount of records for each layer or protocol. As expected, \textbf{Dataset~1} yields more records than \textbf{Dataset~2} due to its size.

\subsection{Experiments}
As noted before, some layers and protocols produce higher amounts of records with different number of fields and length. Consequently, their corresponding stages are more time consuming and memory hungry. Thus, the first experiment to be considered is the sequential execution of each analysis stage for each dataset in order to find out which stages require more  memory or processing time. 

These results serve to drive conclusions about how some stages can be parallelized in order to optimize both time and memory usage for different scenarios. Therefore, the second experiment involves testing different stage scheduling for the optimization of memory usage and execution time, looking for a compromise between both.
Decreasing the execution time is essential to provide a report that covers the daily behavior of the IT infrastructure, within a reasonable execution time, and then roll out the corrective actions, should they be needed.

We set our objective to the goal of processing the corresponding enriched records from 1~TByte worth of traces in less than a quarter of a day, because the traffic volume from large data centers may reach some TByte per day on average as seen in the \textbf{Dataset 2}.

\section{Prototype system evaluation}
\label{sec:eval}
 In this section, we present the performance evaluation of a prototype implementation of an automatic traffic reporting system, which features several sections, such as topology; MAC, IP, UDP conversations; TCP connections; HTTP, DNS transactions, ICMP messages, among others. The report has  more than 75 tables and 48 figures, with about 10 tables and charts per section. Such prototype system was tested on an Intel~\textregistered~Xeon E5-2640 v3 @ 2.6Ghz host with 128GB of RAM and two RAID 0 storages of 16TB each that provides read and writing speeds up to 10 Gbit/s, ideal for traffic sniffing at multi-Gigabit-per-second rates. This set-up is similar to those that can be found in datacenter network probes.

A single RAID system was used to store the enriched records for the analysis. We note that our system, as many other network analysis systems, is favored by the vertical scalability. Thus, a faster processor would provide much better performance.

\subsection{Hindrance finding}
Sequential execution of the system confirmed (see Figure~\ref{fig:stage_times}) that some sections, such as TCP, are prone to require more time owing to the complexity of the metrics. As Figure~\ref{fig:schedulings} shows, the sections have different memory usages as well. All of this might lead to some form of stage starvation in which shorter stages await for a very long time, instead of being processed in parallel along with the most resource consuming tasks or prior to them.
These results helped us to identify certain sections of the report generation that could be delivered earlier if they were executed with a better scheduling. \looseness=-1 

\subsection{Better scheduling}

In this light, we conducted tests for each dataset. In the first place, we used a scheduling in which all stages run in parallel. Afterwards, we tried a hand-crafted smart scheduling which gathers the least resource hungry stages in one process and queues the longest (such as burst and TCP stages) in another process, being both processes executed in parallel. These approaches are compared to the sequential scheduling in Figure~\ref{fig:schedulings}. As shown, execution times are halved in both parallel and smart scheduling. However, memory usage is very intensive for the former, while the latter provides a good compromise in both time and memory usage. Furthermore, we note that the very same behavior can be observed for both datasets. Therefore, we note that smart ordering of execution of the different report sections is cornerstone to achieve a good performance.

\section{Conclusions}
\label{sec:conclusions}
In this paper, we have shown the challenges of automatic traffic reporting systems with large data volumes, which provide the final step of transforming flow records into a human-readable report. These challenges range from the low-level I/O routines performance, to the selection of KPIs and data visualization. In this paper, we have provided a number of recommendations and design tips that are most valuable to other researchers and practitioners in the field for the efficient design of automatic traffic reporting systems. \looseness=-1 

More specifically, we have shown that the use of a high-level language, complemented by low-level I/O routines, provides ample functionality and enough performance to cope with large volumes of data in commodity hardware, as shown in the trace-driven performance evaluation with real-world traffic traces, FERMIN has exceeded our initial performance goals, being able to analyze significant volumes of data in a timely manner.


\section*{Acknowledgments}
This work has been partially supported by the Spanish Ministry of Economy and Competitiveness and the European Regional Development Fund under the projects TR\'AFICA (MINECO/FEDER TEC2015-69417-C2-1-R) and Procesado Inteligente de Tr\'afico (MINECO/FEDER TEC2015-69417-C2-2-R). The authors also thank the Spanish Ministry of Education, Culture and Sports for a collaboration grant.

\bibliography{bibliography}%

\clearpage

\section*{Author Biography}

\begin{biography}{\includegraphics[width=1in,height=1in,clip,keepaspectratio]{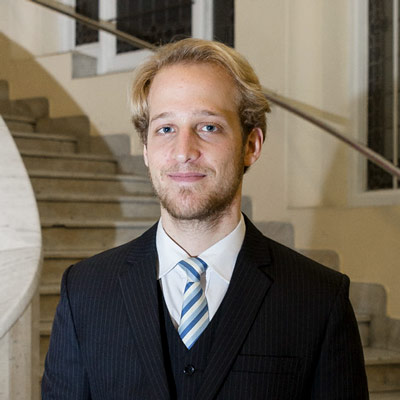}}{\textbf{Carlos G. Vega Moreno} received his M.Sc and B.Sc. degree in Computer Science from Universidad Aut\'onoma de Madrid, Spain, in 2014. He joined the High Performance Computing and Networking Research Group at the same university as a researcher in the Network of Excellence InterNet Science, where he collaborated on European research projects. His current research topics as a Ph.D candidate include log collection and network traffic analysis.}
\hfill \href{mailto:carlosgonzalo.vega@predoc.uam.es}{carlosgonzalo.vega@predoc.uam.es}
\end{biography}

\begin{biography}{\includegraphics[width=1in,clip,keepaspectratio]{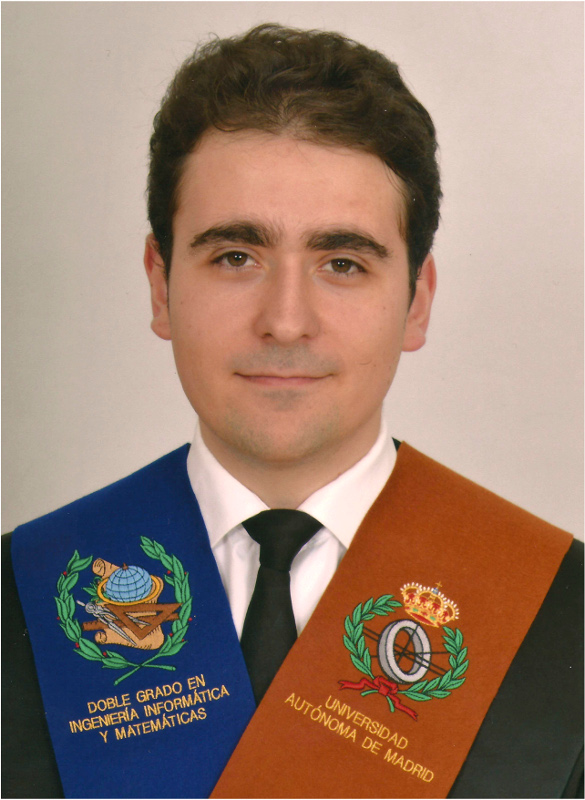}}{{Eduardo Miravalls Sierra} received his double B.Sc. degree in Computer Engineering and Mathematics in 2016 and his M.Sc in Computer Science in 2017 from Universidad Aut\'onoma de Madrid, Spain. He joined the High Performance Computing and Networking Research Group at the same university as an intern in 2015 where he collaborated on the European research project Fed4Fire and the Racing Drones national research project. Right now is working at GMV in GNSS projects.}
\hfill \href{mailto:eduardo.miravalls@uam.es}{eduardo.miravalls@uam.es}
\end{biography}

\vspace{0.5cm}

\begin{biography}{\includegraphics[width=1in,clip,keepaspectratio]{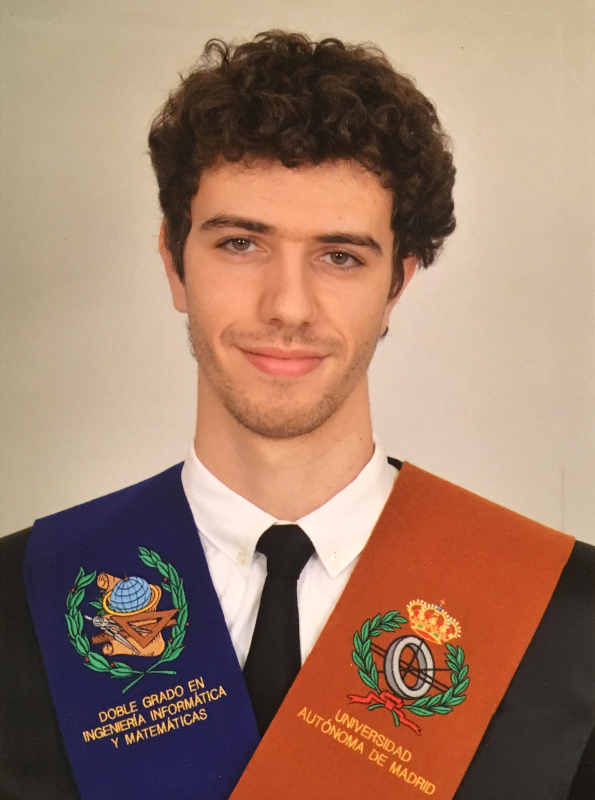}}{{Guillermo Juli\'an Moreno} received his double B.Sc. degree in Computer Engineering and Mathematics from Universidad Aut\'onoma de Madrid, Spain, in 2016, and is currently finishing a M.Sc. in Computational Science and Engineering at \'Ecole Polytechnique F\'ed\'erale de Lausanne, Switzerland. He has been collaborating in the High Performance Computing and Networking research group since 2013, working in network capture drivers and network analysis.}
\hfill \href{mailto:guillermo.julian@estudiante.uam.es}{guillermo.julian@estudiante.uam.es}
\end{biography}

\vspace{0.5cm}

\begin{biography}{\includegraphics[width=1in,height=1in,clip,keepaspectratio]{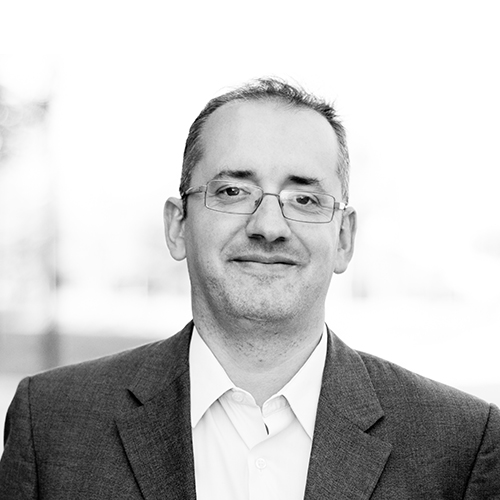}}{{Jorge L\'opez de Vergara}
is associate professor at Universidad Aut\'onoma de Madrid (Spain), and founding partner of Naudit HPCN, a spin-off company devoted to high performance traffic monitoring and analysis. He received his MSc and PhD degrees in Telecommunication Engineering from Universidad Polit\'ecnica de Madrid (Spain) in 1998 and 2003, respectively. He researches on network and service management and monitoring, having co-authored more than 100 scientific papers about this topic.}
\hfill \href{mailto:jorge.lopez_vergara@uam.es}{jorge.lopez\_vergara@uam.es}
\end{biography}

\begin{biography}{\includegraphics[width=1in,height=1in,clip,keepaspectratio]{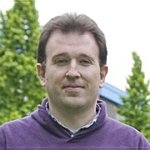}}{{Eduardo Maga\~na Lizarrondo}
received his M.Sc. and Ph.D. degrees in Telecommunications Engineering from Public University of Navarra (UPNA) (Spain), in 1998 and 2001, respectively. He is an associate professor at UPNA. During 2002 he was a postdoctoral visiting research fellow at the Department of Electrical Engineering and Computer Science, University of California, Berkeley. His research interests are network monitoring, traffic analysis and performance evaluation of communication networks.}
\hfill \href{mailto:eduardo.magana@unavarra.es}{eduardo.magana@unavarra.es}
\end{biography}

\begin{biography}{\includegraphics[width=1in,height=1in,clip,keepaspectratio]{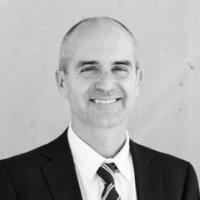}}{{Javier Aracil Rico}
received the M.Sc. and Ph.D. degrees (Honors) from Technical University of Madrid in 1993 and 1995, both in Telecommunications Engineering. In 1995 he was awarded with a Fulbright scholarship and was appointed as a Postdoctoral Researcher of the Department of Electrical Engineering and Computer Sciences, University of California, Berkeley. In 1998 he was a research scholar at the Center for Advanced Telecommunications, Systems and Services of The University of Texas at Dallas. He has been an associate professor for University of Cantabria and Public University of Navarra and he is currently a full professor at Universidad Aut\'onoma de Madrid, Madrid, Spain. His research interest are in optical networks and performance evaluation of communication networks. He has authored more than 100 papers in international conferences and journals. \looseness=-1 } 
\hfill \href{mailto:javier.aracil@uam.es}{javier.aracil@uam.es}
\end{biography}

\end{document}